# Sound velocity measurement methods for porous sandstone. Measurements, finite element modelling, and diffraction correction

Mathias Sæther [1], Per Lunde [1,3], Geir Ersland[2]

[1] Acoustics group, Department of Physics and Technology, University of Bergen, Postboks 7803, N-5020 BERGEN, Norway
[2] Petroleum and process technology group, Department of Physics and Technology, University of Bergen, Postboks 7803, N-5020 BERGEN, Norway
[3] Christian Michelsen Research AS (CMR), P.O. Box 6031 Postterminalen, N-5892 BERGEN, Norway
Contact email: mathias.sather@uib.no

## Abstract

Acoustic material parameters of gas hydrate bearing porous rocks are important for evaluation of methods to exploit the vast methane gas resources present in the earth's subsurface, potentially combined with CO2 injection. A solid buffer method for measuring changes of the compressional wave velocity in porous rocks with changing methane hydrate contents under high-pressure hydrate-forming conditions, is tested and evaluated with respect to effects influencing on the measurement accuracy. The limited space available in the pressure chamber represents a challenge for the measurement method. Several effects affect the measured compressional wave velocity, such as interference from sidewall reflections, diffraction effects, the amount of torque (force) used to achieve acoustic coupling, and water draining of the water-saturated rock specimen. Test measurements using the solid buffer method in the pressure chamber at atmospheric conditions are compared to independent measurements using a water-bath immersion measurement method. Compressional wave velocity measurements have been done in the steady state region at frequency 500 kHz for various specimen made of plexiglas and Bentheim sandstone. Finite element simulations of the solid buffer measurement method with plexiglas specimen have been used for comparison with the measurements, and to aid in the design, control, and evaluation of the measurement method and results. Highly favorable agreement between the two measurement methods has been obtained, also with respect to repeatability and reproducibility. The results indicate that the solid buffer method may be suitable for use in the pressure chamber with Bentheim sandstone and changing methane hydrate contents under high-pressure hydrate-forming conditions, for quantitative measurements of the compressional wave velocity in such rock core samples at these frequencies.

## 1 Introduction

There is an increasing interest in exploiting the methane gas resources present in natural gas in the earth's subsurface. Even if conservative estimates are considered, the consensus is that hydrocarbon gas hydrate resources are vast [1–4].

Given the more stable nature of $CO_2$ - containing hydrates, safe storage of $CO_2$ might be accomplished in porous rock resorvairs. This has already been tested by injecting $CO_2$ from the Sleipner oil and gas field into the Utsira formation. Potential leakage and other safety concerns have been studied in detail by [5–7]

Gas hydrate or ice in sediment pores will significantly affect the acoustic parameters such as compressional and shear wave velocities, $c_L$, $c_S$, and compressional and shear wave absorption coefficients, $\alpha_L$, $\alpha_S$) [8, 9]. This can potentially be used to monitor the amount of gas hydrates in porous rocks, and has been indicated to be the most promising method to detect hydrate deposits remotely [1]. Acoustic parameters also play and important role in reservoir simulations [10] and as input parameters in the Biot model [11, 12]. These models can give insight in physical and geological properties of the porous rock and the hydrates trapped in its pores.

There are numerous publications on wave velocity measurements in rocks. In [13–17] pulse methods were used to measure compressional and shear wave velocities at different pressures and temperatures. In these works, transducers were attached directly to the specimen, and the setups were immersed in pressurized oil cells inducing a hydrostatic pressure. It has been pointed out that the oil may penetrate some of the pores in the rocks that may influence on the results [18].

Biot [11, 12] developed a theory for propagation of elastic waves in a fluid-saturated porous solid. Dispersion curves for for phase velocities, group velocities and attenuation factors were presented. Input parameters in the model are bulk modulus and mass density of the sand grains, permeability, turtuosity, and porosity of the sediment, viscosity, bulk modulus and mass density of the fluid and shear and bulk moduli of the sediment frame.

In [18] McSkimin categorizes and discusses strengths and weaknesses of different acoustic measurement methods for various media ranging from low and high viscosity fluids to rubbers, plastics and metals. The two main categories for measuring attenuation and velocities in solids are divided into pulse and resonance techniques. These two techniques can be used with e.g the immersion method and the solid buffer method. Many authors make use of transit time measurements of a pulse when finding wave velocities. This is typically done by taking the first rise of the pulse and correct for time delays in the electronics [13–17]. In measurements where unwanted reflections may occur this pulse method is easy to use. However, the accuracy of this way of finding time of flight is debated e.g in [19–21]. For highly attenuating media, Futterman [19] found a logarithmic behavior at the rise of the signal and concludes that the very onset of the signal would be hard to pinpoint. Molyneux et al. [20, 21] also discuss how high frequency components will be overrepresented at the rise of the signal, making precise time of flight measurements and dispersion measurements hard to do.

McSkimin [18] outlines how measurements with the pulse method can be used to find acoustic wave velocities and attenuation for distinct frequencies. Winkler and Plona [22] developed a method using Fourier spectroscopy of short pulses to measure dispersion of compressional wave velocity in rocks based on the work of of Papadakis [23], Sachse and Pao [24]. The short pulses needed in this method are easy to separate in time domain. The results were corrected for diffraction effects. This method uses short pulses and might be less prone to unwanted reflections. Several authors have used this method for dispersion measurements in solids [22, 25–27].

Several labratory studies on acoustic wave velocities in methane-hydrate bearing sediments have been done as methane hydrate research is becoming increasingly relevant. Helgerud et. al [28] presented two different effective medium models for elastic waves in unconsolidated ocean bottom sediments based on Gassmann's equation [29]. In the first model it was assumed that hydrate is a component of the solid frame without affecting the pore fluid. In the second model they assumed that hydrate is modifying the elastic properties of the pore fluid but not the solid frame. The results were found to qualitatively agree with vertical sismic profiling (VSP) data logs. Later compressional and shear wave velocities were mesured in polycrystalline ice and pure gas hydrate specimen [30]. A 1 MHz shear wave transducer was used to measure both compressional and shear waves. The transducer was excited with short pulses and the first rise of the signal was used to detect the transit time.

Winters et al. [8] used a similar approach as Helgerud et al. in modeling of hydrates in sediments. They also measured compressional wave velocities on hydrate bearing sediments from different sites and from specimen formed in the laboratory. From this it was indicated that hydrates contribute differently to the elastic moduli in natural sediments and in specimen made in the labaratory. Later Winters et. al [31] measured compressional wave velocities in coarse-grained sediment for different pore space occupants. The compressional wave velocity in water-saturated specimen was found to be substantially lower than in hydrate-bearing specimen. The velocity measured in fine-grained sediments containing hydrates were reported to be substantially lower than in the coarse-grained sediments. In both these studies the compressional wave velocity was measured by exciting a 1 MHz transducer with short pulses and finding the first rise of the signal.

Chand et al. [32] compared different models with available seismic data. Four different models, the "weighted equation" (WE) [33], "three-phase effective-medium theory" (TPEM) [34], "three-phase Biot theory" (TPB) [35, 36] and "differential effective-medium theory" (DEM) [37–39] were compared. In general the four models estimate wave velocities that are consistent for fluid saturated sediments, but differs when hydrate is introduced in the models.

Waite et al. [9] measured the compressional wave velocity and studied three ways that methane hydrate can interact with fluid-saturated sediments: Hydrate forming primarily in the pore fluid, hydrate becoming a load-bearing member of the sediment matrix, and hydrate cementing sediment grains. They found that in partially water-saturated, methane-hydrate bearing Ottawa sands, hydrate is likely surrounding the grains and cementing them. Waite et al. used the same measurement setup and measurement methods as Winters et al. [8].

Rydzy et al. [40] measured the compressional wave velocity during hydrate growth in unconsolidated sand in the gas hydrate stability zone. It was found that the wave velocity continued to increase dramatically for nine hours after a change in velocity first was observed. Measurements were also done with increasing confining pressure after hydrate forming was finished. 100 kHz transducers were used with short pulses and time detection in the same manner as in [8, 9, 30].

Hu et al. used bender elements to measure shear and compressional waves in consolidated and unconsolidated sediments during hydrate growth and hydrate dissociation [41]. They aimed at finding a relation between gas hydrate saturation and acoustic velocities of the hydrate-bearing sediments. The measurements were interpreted with models used in [32]. It was found that acoustic velocities differ for the same degree of hydrate saturation during the hydrate-formation process and the hydrate-dissociation process. This was explained by the fact that hydrate will affect the pore fluid, sediment frame and sediment grains differently in the two processes. The wave velocities were found using a Fast Fourier Transform (FFT) method, and are reported to increase rapidly for hydrate saturations 10 -30 %.

In the present project, the long term objective is to monitor the compressional and shear wave velocities, and the compressional and shear wave absorbtion coefficients in a Bentheim sandstone during methane hydrate growth in a pressure cell. The objective in the present paper is to test and evaluate a solid buffer method that is to be used in this pressure cell for compressional wave velocity measurements. Because of the limited lateral space inside the cell, unwanted reflections are present. The impact of these reflections at atmospheric conditions are to be adressed before carrying out with hydrate growth experiments under pressure. Even short pulses that are needed in the Fourier spectrum method may suffer from side reflections and thus be unseperable and subject to interference in the time domain.

To test and evaluate the solid buffer method implementation in the pressure cell measurement setup, an independent measurement method has been used. A water bath combined with an immersion method [18] has been used for this purpose, for comparison with the solid buffer method and evaluation of the results. Measuring with the immersion method is assumed to give results with no unwanted reflections present. Moreover, in the solid buffer method, an applied force is needed to keep transducers, buffers and specimen pressed together for proper contact. As pointed out in [13–17], the wave velocity will be affected by the applied force. The impact of this applied force has been investigated in comparison with the immersion method.

Finite element modeling of the solid buffer method has also been used for comparison with measurements and to aid in the control and evaluation of the measurement method and results. Bentheim sandstone is a highly attenuating, non-homogenous material which compressional wave velocity is hard to simulate in a finite element modeling tool. The fluid-saturation might change over time and the amount of pressure excerted on the rocks to have proper contact may lead to poor measurement reproducibility. For these reasons, test measurements on plexiglas specimen with approximately the same dimensions and compressional wave velocity as the Bentheim specimen have also been done.

All solid buffer measurements shown in this paper, have been done in the pressure cell, operated in air at 1 atm. and in room temperature conditions.

The present paper represents a further analysis of measurement methods discussed in [42]. Even if compressional wave velocity obtained with Fourier spectrum analysis and time of flight measurements gave consistant results in [42], further investigation on how unwanted reflections and diffraction effects influence in the present solid buffer measurement method to be used in the pressure cell.

## 2 Experimental setup and measurement methods

### 2.1 Solid buffer method

In the solid buffer method two different measurements are used (Figure 1): Measurement A without specimen inserted, and measurement B with the specimen inserted into the wave propagation path. The compressional wave velocity in the specimen is found by taking the difference between the transit time in measurement A and the transit time in measurement B. The specimen are a plexiglas disc with 20 mm thickness and 50 mm diameter, a plexiglas cylinder with 60 mm thickness and 50 mm diameter, a Bentheim sandstone disc with 20 mm thickness and 50 mm diameter, and two Bentheim sandstone cylinders with 50 mm diameter and thicknesses 47 mm and 53 mm, respectively.

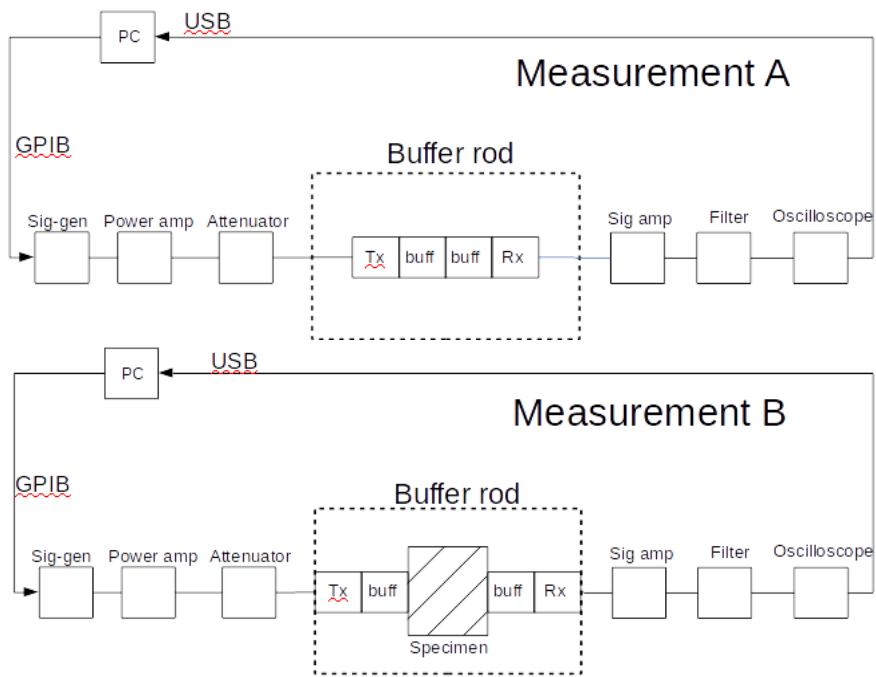

**Figure 1:** Schematics of the solid buffer method.

In the measurements, two in-house constructed 500 kHz transdcers are used. The transmitting transducer is driven by an Agilent 33250A signal generator connected to an ENI 240L RF power ampflier and a Hatfield Attenuator 2105. A 15 periods, 500 kHz sinusoidal signal is used. The receiving transducer is connected to a Panametrics Ultrasonic Preamp, a HP filter, and a Tektronix DPO 3012 oscilloscope. The filter is used to ensure that low frequency components in the signal is attenuated so that a steady-state region can be achieved. The signal generator is controlled by a PC via GPIB communication and the signal is aquired with the oscilloscope via an USB connection. All solid buffer measurements are done at atmospheric conditions in the pressure cell. The available space inside the pressure cell is a tube (outer sleeve in Figure 3) with inner diameter 50 mm and length 30 cm. In the next phase of the project, methane gas will be flowing inside the pressure cell and the porous rock needs to have a certain surface exposed to the gas for the gas to efficiently seep into the porous rock. The diameter of the constructed transducers and the buffers is thus 35 mm. The total length the transducers, buffers and the specimen can maximum be 30 cm. Specimen with length up to 10 cm is of interest and thus leaving 10 cm left for transducers and buffers on both sides. A transducer with proper backing will here be 5 cm long, which leaves 5 cm for the buffer. A plexiglas buffer of this size, with a compressional wave velocity of 2700 m/s will have room for 18 periods before reflections from the end of the buffers arrive, which should be sufficient in these measurements.

To obtain reproducible measurements, the transducers, buffers and specimen need to be aligned and pressed together with same force in every measurement. The buffers are glued directly to the transducers permanently with epoxy, and a coupling fluid is applied on the buffer/buffer surface or the buffer/specimen surface. The transducers are centralized by placing them in mounting cups, and a torque key is used on the end caps to exert 1 Nm torque on the mounting cups and transducers in all measurements, see Figure 2.

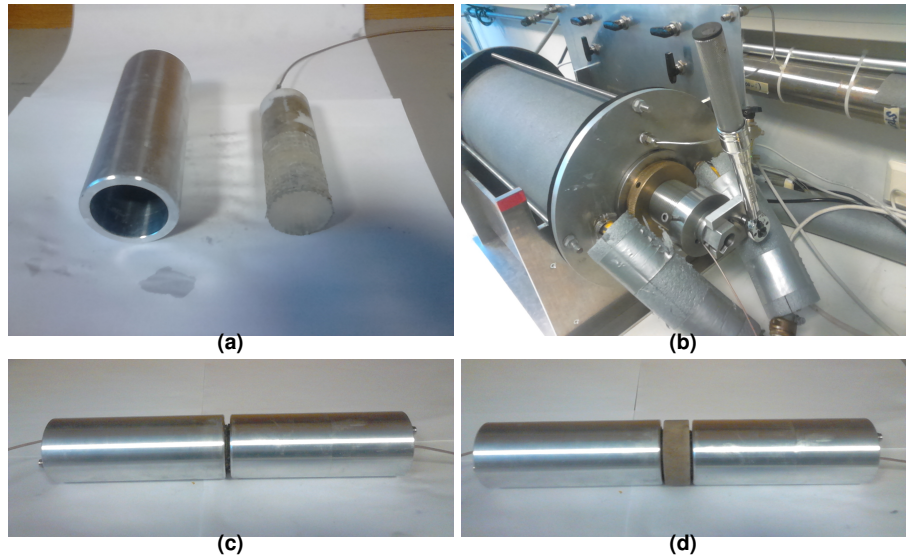

**Figure 2:** Photograps of **a)** mounting cup and transducer, **b)** pressure cell with torque key, **c)** transducers with mounting cups, **d)** transducers with mounting cups and Bentheim sandstone specimen.

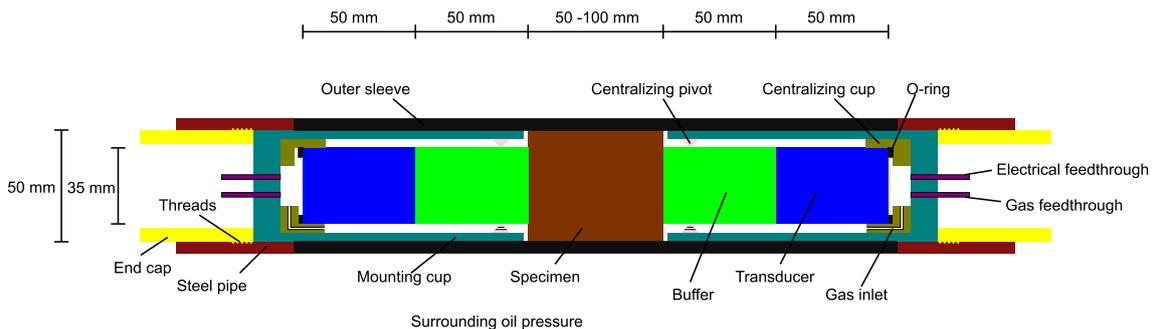

**Figure 3:** Schematics of the solid buffer measurement setup in the pressure cell.

The transit transit time measurements in measurement A and B, respectively, $t^A$ and $t^B$, are decomposed as

$$\begin{aligned}
t^A &= t_{gen} + t_{eltr1} + t_{b1}^{p,A} + t_{b2}^{p,A} + t_{eltr2}^{A} - t^{dif,A}, \\
t^B &= t_{gen} + t_{eltr1} + t_{b1}^{p,B} + t_{m}^{p} + t_{b2}^{p,B} + t_{eltr2}^{B} - t^{dif,B},
\end{aligned} \quad (1)$$

where $t_{gen}$ is the time delay in the signal generator, $t_{eltr1}$ and $t_{eltr2}$ are the time delays in the transmitting and receiving electronics and transducers, respectively. $t_{b1}^{p,A}$, $t_{b1}^{p,B}$ and $t_{b2}^{p,A}$, $t_{b2}^{p,B}$ are the plane wave transit times in the buffers at the transmitting and receiving sides, respectively, for measurement A and B. $t_m^p$ is the plane wave transit time in the specimen, $t^{dif,A}$ and $t^{dif,B}$ are the diffraction correction terms in measurements A and B, respectively. It is assumed that $t_{b1}^{p,A} = t_{b1}^{p,B}$, $t_{b2}^{p,A} = t_{b2}^{p,B}$, $t_{eltr2}^{p,A} = t_{eltr2}^{p,B}$. Since $t_m^p = M/c_L$, where $M$ and $c_L$ are the length and compressional wave velocity of the specimen, $c_L$ is obtained as [18, 43],

$$c_L = \frac{M}{t^B - t^A + (t^{dif,A} - t^{dif,B})} \tag{2}$$

The time delays due to diffraction correction are given as [43]

$$\begin{aligned} t^{dif,A} &= -\frac{\angle H^{dif,A}}{\omega}, \\ t^{dif,B} &= -\frac{\angle H^{dif,B}}{\omega}, \end{aligned} \tag{3}$$

where $\omega = 2\pi f$ is the angular frequency, $f$ is the frequency. $H^{dif,A}$ and $H^{dif,B}$ are the diffraction corrections of measurement A and B, respectively, here approximated with diffraction correction for a uniformly vibrating circular and plane piston mounted in a rigid baffle of infinite extent (referred to as the "baffled piston diffraction correction", BPDC). The diffraction corrections are then given as [44, 45]

$$H^{dif,A} = \frac{\langle p^A \rangle}{p^{plane,A}}, \qquad H^{dif,B} = \frac{\langle p^B \rangle}{p^{plane,B}}, \tag{4}$$

where $\langle p^A \rangle$ and $\langle p^B \rangle$ are the average freefield pressures at the receiver front surface in measurements A and B respectively, and $p^{plane,A}$ and $p^{plane,B}$ are the plane wave pressures of the piston filed at the center of the receiver front surface in measurements A and B, respectively. BPDC applies to an infinite medium and was developed for waves propagating in fluids [44, 45]. For $ka >> 1$ Rogers and Van Buren [46] derived a simplified expression for BPDC that is independant of $ka$ when $k$ is the wavenumber and $a$ is the radius of the piston. In this work the radius of the piezoelectric element has been used as $a$. For the additional assumption that the media can be described as fluids, Rogers and Van Buren's expression has also been used for waves propagating through multiple media [23, 25, 46–48]. With the Fresnel condition satisfied ($k_z = k - \frac{k_r^2}{2k}$ where $k_r$ and $k_z$ are spatial and axial wave numbers), only the waves propagating at small angles to the axis contribute to the total field [48]. This means that when calculating the diffraction correction, a total Fresnel parameter can be used, consisting of the sum of the Fresnel parameters of the media in the wave propagating path. Assuming that the BPDC can be used in solids and $ka >> 1$, the diffraction corrections are here approximated by [23, 25, 46–48]

$$H^{dif,A} = 1 - e^{-i(2\pi/S_y)}\left(J_0(2\pi/S_y) + iJ_1(2\pi/S_y)\right), \qquad S_y = \frac{Yc_y}{fa^2}, \tag{5}$$

$$H^{dif,B} = 1 - e^{-i(2\pi/S_{ym})}\left(J_0(2\pi/S_{ym}) + iJ_1(2\pi/S_{ym})\right), \qquad S_{ym} = \frac{Yc_y}{fa^2} + \frac{Mc_L}{fa^2}, \tag{6}$$

were $J_0$ and $J_1$ are the zeroth and first order Bessel functions of the first kind. $S_{ym}$ and $S_y$ are the Fresnel parameters at the receiver surface in measurement A and B, respectively. $Y$ is the total length of the two buffers, and $M$ is the length of the specimen. $c_y$ and $c_L$ are the compressional wave velocities in the buffers and the specimen, respectively.

When finding $(t_B - t_A)$ in Eqs. 2, different features of the signal can be used. Many authors have used the first arrival of the signal [15], [16], [17], or a Fourier spectrum method, [22, 25–27]. Here we use the time difference of zerocrosses in the signal and defined it as the zerocross technique. Zerocross number 1 means the earliest zerocross detected in measurement A or measurement B (the single zerocross with no averaging, indexed 1 in Figure 4). Compressional wave velocities are presented here with the time difference between zerocross number 1,2,3... and zerocross number 1,2,3 in measurement A and B, respectively. The Fourier spectrum technique has been compared with the zerocross technique in [42]. Here, zerocrosses have been used to find $(t_B - t_A)$ because they give more information about when ventual unwanted reflections interfere with the signal.

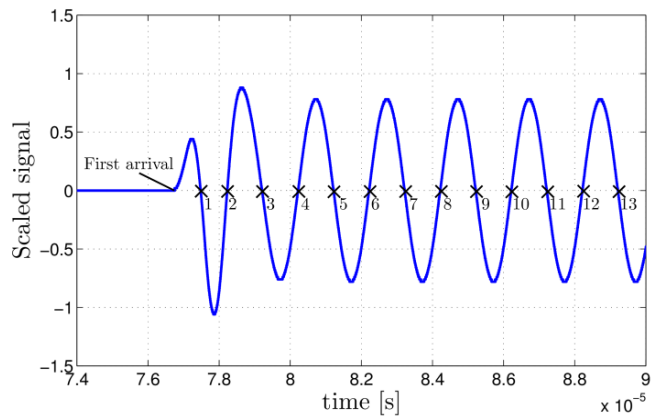

**Figure 4:** Definitions of zerocross number and first arrival.

## 2.2  Immersion method

In the immersion method it is assumed that no unwanted reflections are present and the method serves to be suitable as a reference measurement method for comparison with the solid buffer method. As for the solid buffer method, a transit time measurement A and a transit time measurement B are used. These transit time measurements are used to calculate the compressional wave velocity in the immersed specimen. When immersing the specimen in the water bath, an adapter is used to keep it steadily in-between the two transducers. The specimen are the same as used with the solid buffer method.

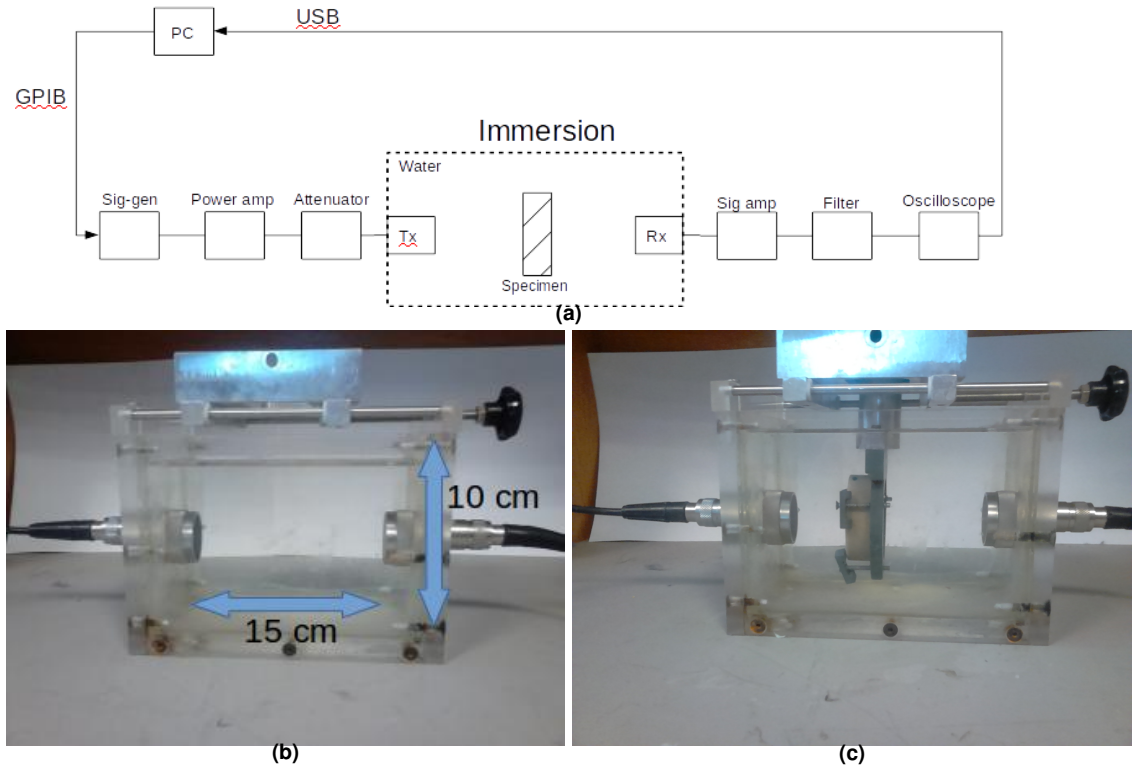

**Figure 5: a)** Schemeatics of the experimental setup of the immersion method, **b)** water immersion cell, measurement A, **c)** water immersion cell, measurement B.

In the measurements, two commercial transducers (Olympus Panametrics-NDT- V302 1.0 MHz ) are used in the water immersion cell showed in Figure 5. The outer diameter of the transducers are 31.5 mm. The piezoelectric element is assumed to have a radius of 12.5 mm which is the same size of the piezoelectric element in the relevant immersion transducers found on the Olympus homepages, [49]. The measurement setup is the same as in the solid buffer method except for using the Olympus transducers and the immersion cell instead of the transducers with buffers. A 15 periods, 500 kHz sinusoidal signal is used. The water immersion cell has height 10 cm, width 10 cm and length 15 cm. The distance between the transducers is 11 cm. With no unwanted reflections, the measured transit times are decomposed as

$$\begin{aligned} t^A &= t_{gen} + t_{eltr1} + t^p_{w1} + t^p_{w2} + t^p_{w3} + t^A_{eltr2} - t^{dif,A}, \\ t^B &= t_{gen} + t_{eltr1} + t^p_{w1} + t^p_m + t^p_{w3} + t_{eltr2,B} - t^{dif,B}, \end{aligned} \quad (7)$$

where $t^p_{w1}$ and $t^p_{w3}$ are the plane wave transit times in water before and after the specimen, $t^p_m$ is the plane wave transit time in the specimen, and $t^p_{w2}$ is the plane wave transit time in the water region of measurement A that is occupied by the specimen in measurement B. Since $t^p_m = M/c_L$, Eq. 7 gives [25]

$$c_L = \frac{M}{M/c_w + t^B - t^A + t^{dif,A} - t^{dif,B}}, \quad (8)$$

where $c_w$ is the sound velocity in water. The time delays due to diffraction are given as

$$t^{dif,A} = -\frac{\angle H^{dif,A}}{\omega}$$
$$t^{dif,B} = -\frac{\angle H^{dif,B}}{\omega} \quad (9)$$

where the diffraction corrections are approximated by [25, 47]

$$H^{dif,A} = 1 - e^{-i(2\pi/S_w)}\left(J_0(2\pi/S_w) + iJ_1(2\pi/S_w)\right), \quad S_w = \frac{Dc_w}{fa^2},$$

$$H^{dif,B} = 1 - e^{-i(2\pi/S_{wm})}\left(J_0(2\pi/S_{wm}) + iJ_1(2\pi/S_{wm})\right), \quad S_{wm} = \frac{(D-M)c_w}{fa^2} + \frac{Mc_L}{fa^2},$$

(10)

where D is the distance between the transducers, and $S_w$ and $S_{wm}$ are the Fresnel parameters at the receiver surface in measurement A and B, respectively. *a* is the assumed radius of the piezoelectric element, 12.5 mm.

## 3  Simulation Setup

In this work Comsol Multiphysics version 4.2a has been used for finite element modeling. The transducers input electrical admittance and the signal propagation in the solid buffer method have been modeled with the voltage-to-voltage transfer function from transmitter to receiver. A frequency domain study with the structural mechanics module has been used with triangular meshing elements of 10 elements/wavelength at 500 kHz calculated for 2700 m/s in all domains (this means approx 7 elements/wavelength for the piezoelectric disc at 500 kHz). A frequency domain study in Comsol solves the finite element equations for a range of input frequencies. Comsol provides a full finite element solution for the measurement setup from transmitter to receiver. A time domain solution is obtained in the post-processing of the signal through Fourier synthesis with frequency resolution 3 kHz and sampling frequency 500 MHz. It is also possible to add electrical circuit elements to model electrical load from transmitting and receiving electronics. In this work only transducer(s) and medium have been modeled with Comsol. The effect of a Thevenin circuit has been taken into acount in the post-processing of the signal.

### 3.1  Transducer simulation parameters

The transducer construction is shown in Fig 6, the materials used are indexed and shown in Table 1. Typical compressional and shear wave velocities for use in the simulation study have been measured roughly using the solid buffer method outside the pressure cell. The first rise of the signal has been used for time detection to determine the wave velocities. Transducers Olympus V151 and Olympus Panametrics-NDT-V302 was used. The immersion method with distinct frequencies outlined in [18, 25] was used for attenuation measurents. The experimental setup of this method is the same as in the immersion method presented here, but the measured quantity in measurement A and B is here the amplitude of the voltage signal. The specimen used to determine the material parameters are thin and hence the differences in amplitude due to diffraction are assumed to be small.

The Q values are measured for compressional waves with no diffraction correction and $Q_L = Q_S$ is assumed.

The only value not measured is the Q-value for the POM material. This is set very low to dampen some of the lateral reflections in the transducer. Little change is however seen in the simulations by varying the Q-value of the POM material. The piezoelectric material constants used in the simulations are shown in Table 2 [50]. The buffer is plexiglas and the backing is a mixture with tungsten/epoxy mass-ratio 4:1. The tungsten grains have a diameter of approx. 5 microns. The transducer casing is made of POM and the green regions in the transducer construction are Epofix epoxy.

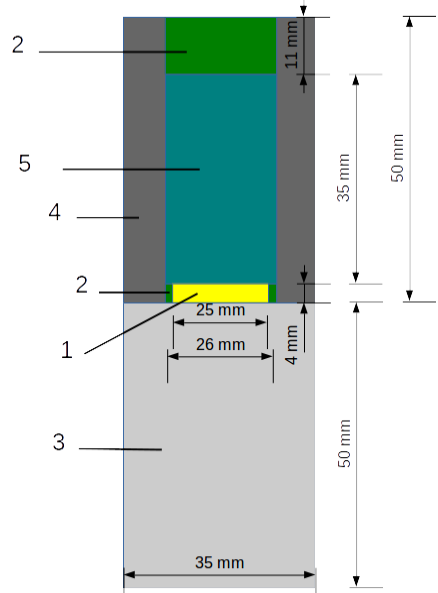

**Figure 6:** Transducer construction.

A quite low $Q$ value for the Epofix is measured. This is probably due to air bubbles trapped inside the epoxy. The geometry of the specimen used to measure the properties in Table 1 was either a circular disc (Epofix and epoxy/tungsten) or a small plate (plexiglas and POM). The densities were measured by calculateing the volume and measuring the mass of the specimen. After setting up the geometry in Comsol with all material parameters in Table 1 and 2, a frequency range, 100-900 kHz with 3 kHz steps, is given as input parameters in the model to model the input electrical impedance of the transducer.

**Table 1:** Material properties used in the simulations. (∗: not measured.)

| Index | Material | $c_L$ [m/s] | $c_S$ [m/s] | $\rho$ [kg/m$^3$] | Q |
|---|---|---|---|---|---|
| 1 | Pz27 disc, 25x4 mm | - | - | - | - |
| 2 | Epofix | 2630 | 1460 | 1120 | 30 |
| 3 | Plexiglas | 2700 | 1398 | 1184 | 63 |
| 4 | POM | 2430 | 1215 | 1420 | 10* |
| 5 | Epoxy/Tungsten | 1695 | 843 | 5500 | 35 |

Table 2 gives material constants for the Pz27 piezoelectric ceramic material. Meggitt [51] gives a set that uses a $Q$-value and a loss-angle, $\delta$ for the mechanical and electrical loss, respectively. The piezoelectrical material set, Adjusted(Lohne/Knappskog/Aanes), used in [52] has complex values to include loss and is also used here.

**Table 2:** Piezoelectrical ceramic material constants for Pz27.

| Variable | Meggitt [51] | Adjusted(Lohne/Knappskog/Aanes) [52] |
| --- | --- | --- |
| $c_{11}^E$ [$10^{10}$ N/m$^2$] | 14.7 | $12.025(1 + i/96)$ |
| $c_{12}^E$ [$10^{10}$ N/m$^2$] | 10.5 | $7.62(1 + i/70)$ |
| $c_{13}^E$ [$10^{10}$ N/m$^2$] | 9.37 | $7.42(1 + i/120)$ |
| $c_{33}^E$ [$10^{10}$ N/m$^2$] | 11.3 | $11.005(1 + i/190)$ |
| $c_{44}^E$ [$10^{10}$ N/m$^2$] | 2.3 | $2.11000(1 + i/75)$ |
| $e_{31}^E$ [ C/m$^2$] | -3.09 | $-5.4(1 - i/166)$ |
| $e_{33}^E$ [ C/m$^2$] | 16 | $17(1 - i/324)$ |
| $e_{15}^E$ [ C/m$^2$] | 11.64 | $11.20(1 - i/200)$ |
| $\epsilon_{11}^E$ [$10^{-9}$ F/m] | 10.005 | $8.11044(1 - i/50)$ |
| $\epsilon_{33}^E$ [$10^{-9}$ F/m] | 8.0927 | $8.14585(1 - i/130)$ |
| $\rho$ [kg/m$^3$] | 7700 | 7700 |
| $Q_m$ | 74 | - |
| $\tan \delta$ | 0.017 | - |

### 3.2 Simulation of solid buffer measurement system

Figure 7 shows the measurement system that has been simulated. The effect of the signal generator is taken into account in the post-processing. The receiving and transmitting electronics are neglected and the oscilloscope electrical load is set to be infinite (open circuit). Under these assumptions a full transmit-receive system has been modeled. In the same way as the impedance simulations, after setting up the geometry in Comsol with all material parameters in Table 1 and 2, a frequency range, 100-900 kHz, with 3 kHz steps, is given as input parameters in the model to model the transfer function $V_2/V_1$. The voltage, $V_1$, is set to 1 V and $V_2$ is found with open circuit conditions. After adding the effect of the Thevenin circuit and using a Fourier synthesis of the voltage signal $V_2(f)$, the time tomain signal $V_2(t)$ has been found in the time region 0 - 33 ms and with time resolution 2 ns.

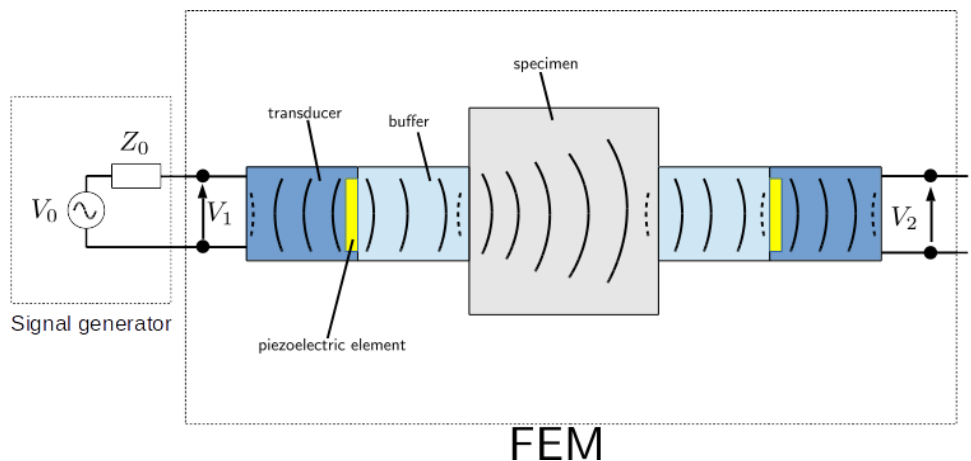

**Figure 7:** Schematic of transducers and specimen simulation configuration.

$V_0(f)$ is the frequency spectrum of the input voltage signal of the signal generator, and $Z_0$ is the internal impedance of the signal generator, here set to 50 Ω. $V_1(f)$ is the frequency spectrum of the input voltage signal to the transducer and $V_2(f)$ is the frequency-spectrum of the output voltage signal at the receiving transducer. The output from Comsol is the transfer function $H_{VV} = \frac{V_2}{V_1}$ but the interesting parameter is $V_2$ for a given signal $V_0$. From the Thevinin circuit given in Figure 8 one has

$$
\begin{aligned}
V_0 &= IZ_0 + IZ_T, \\
V_1 &= IZ_0, \\
\frac{V_1}{V_0} &= \frac{Z_0}{Z_0 + Z_T},
\end{aligned}
\tag{11}
$$

giving

$$
V_2 = V_0 \frac{V_1}{V_0} \frac{V_2}{V_1} = V_0 \frac{Z_0}{Z_0 + Z_T} \frac{V_2}{V_1}.
\tag{12}
$$

The time domain signal of a 15 periods toneburst can now be found by Fourier synthesis of $V_2$.

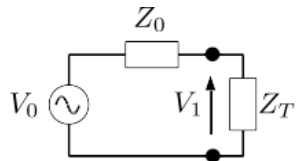

**Figure 8:** Thevinin circuit representing the interaction between the signal generator and the transmitting transducer.

## 4 Transducer construction

In this section the construction of the transducer is described with simulated and measured input electrical conductance plots in Figures 9-10. Consistency with measurements and simulations indicate that the simulations are representative for the real transducer. The active piezoelectric element of this transducer is a 25x4 mm Pz27 disc. Electrical wires are soldered to the electrodes on the flat sides of the piezoelectric element. The POM chassis is glued to the cylindrical surface of the piezoelectric element. Tungsten-epoxy backing material is mixed and poured into the chassis from behind. After the tungsten-epoxy mix has cured, the rest of the space behind the piezoelectric element is filled with Epofix. A plexiglas cylinder with roughened outer surface was used as the solid buffer and glued on to the transducer front using Epofix, see Figure 6.

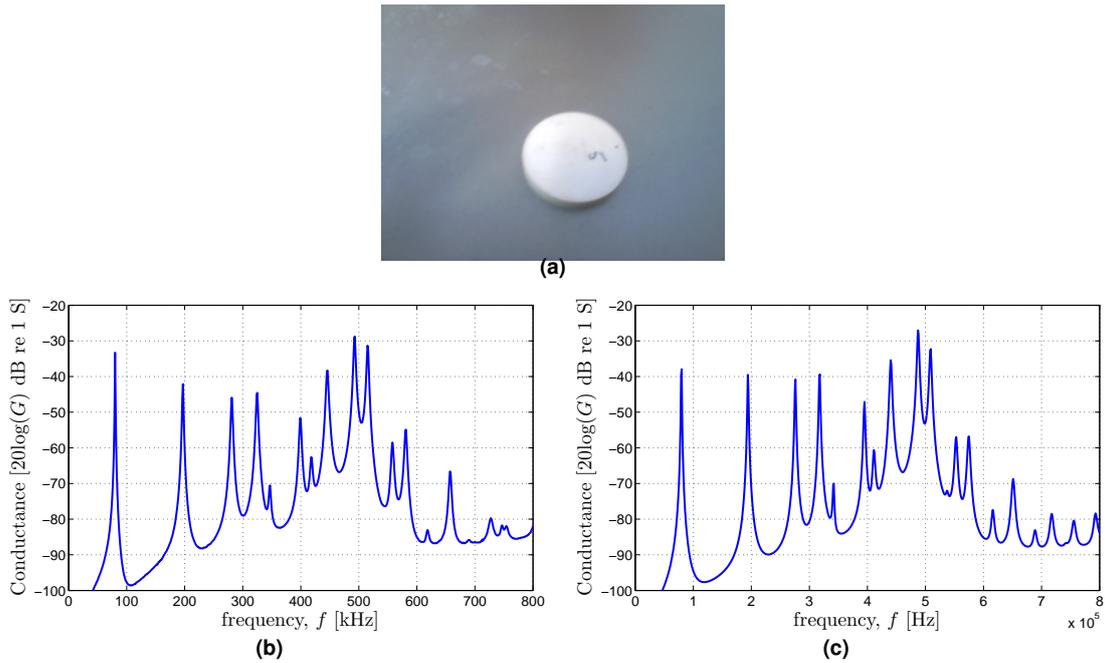

**Figure 9:** Electrical conductance plots for the 25x4 mm piezelectric disc element, Pz27. **a)** Pz27 element. **b)** Measured conductance. **c)** Simulated conductance.

Figure 9 shows the measured and simulated electrical conductance frequency responses of the piezoelectric element. Many of the modes have been successfully simulated, with fair agreement between measurements and simulations. The final transducer with the buffer shown in Figure 10 posesses a highly increased bandwidth compared with the single piezoelectric element in Figure 9. This is seen in both mesurements and simulations of the electrical conductance. However there are more ripples in the simulated conductance than in the measured conductance. This may be due to inaccurate material properties. Especially the tungsten-epoxy backing layer has a large impact on the bandwidth of the transducer and damping of unwanted internal reverbarations.

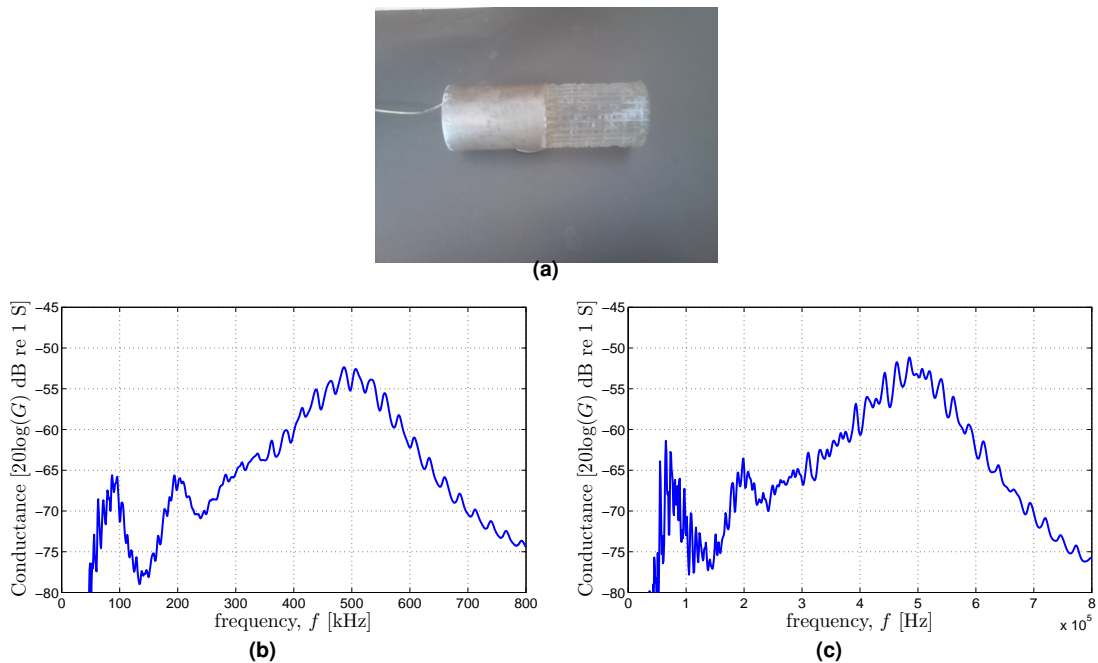

**Figure 10:** Electrical conductance plots for the full transducer with buffer. **a)** Full transducer. **b)** Measured conductance. **c)** Simulated conductance.

## 5 Results

Five different specimen manufactured to fit inside the pressure cell are shown in Figure 11. Two plexiglas specimen, one disc with 20 mm thickness and diameter 50 mm, and one cylinder with 60 mm length and diameter 50 mm were cut out of the same chunck of plexiglas. Two Bentheim porous rock specimen, one disc with 20 mm thickness and diameter 50 mm, and one cylinder with 47 mm length and diameter 50 mm were cut out of the same Bentheim core sample. These two Bentheim specimen have an initial water saturation of 58 %. The third Bentheim porous rock specimen is a cylinder with 53 mm length and diameter 50 mm having an initial water saturation of 62 %. To preserve the water saturation, the Bentheim specimen are kept in a water bath.

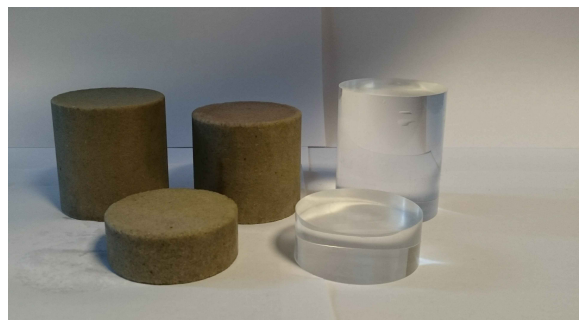

**Figure 11:** Two plexiglas and three Bentheim sandstone specimen used in this study.

### 5.1 Compressional wave velocity, plexiglas specimen.

Figure 12 shows the compressional wave velocity for the two different plexiglas specimen measured using the immersion method. A 15 periods 500 kHz, 100 mV input signal is used on the signal generator. The receiving signal amplifier is set to 60 dB amplification. The red and blue lines show the results with and without diffraction correction applied. In these measurements no or very little side reflections are assumed. After zerocross number 5 the compressional wave velocity reaches a steady state at around 2734-2735 m/s. These two specimen are cut out of the same chunk of plexiglas so the acoustic properties are assumed to be very similar in both these two specimen. It is seen from the figure that the diffraction correction is quite the same for measurements on both the 20 and the 60 mm specimen. The wavelength in this plexiglas is computed to be 5.5 mm which means that internal reflections in the axial direction in the 20 mm plexiglas specimen will occur after 7 periods (zerocross nr 14).

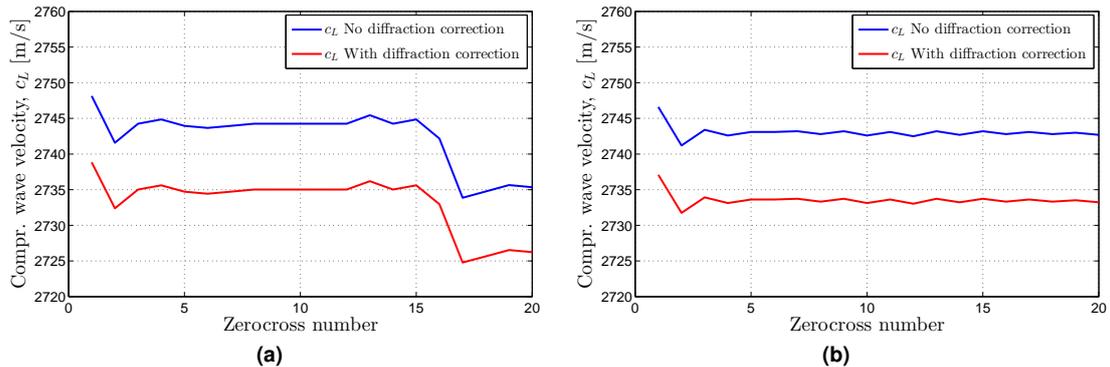

**Figure 12:** Compressional wave velocities for plexiglas specimen measured with the immersion method for the **a)** 20 mm, and **b)** 60 mm specimen.

Measurements on the same plexiglas specimen using the solid buffer method are shown in Figure 13. The results are quite consistent with the immersion method results of Figure 12. The applied torque in these measurements were 1 Nm in both measurement A and B. At the 10th zerocross the compressional wave velocity is 2734 m/s for both specimen. For the 20 mm specimen there is a slight increase in compressional wave velocity which might be due to interference with side reflections.

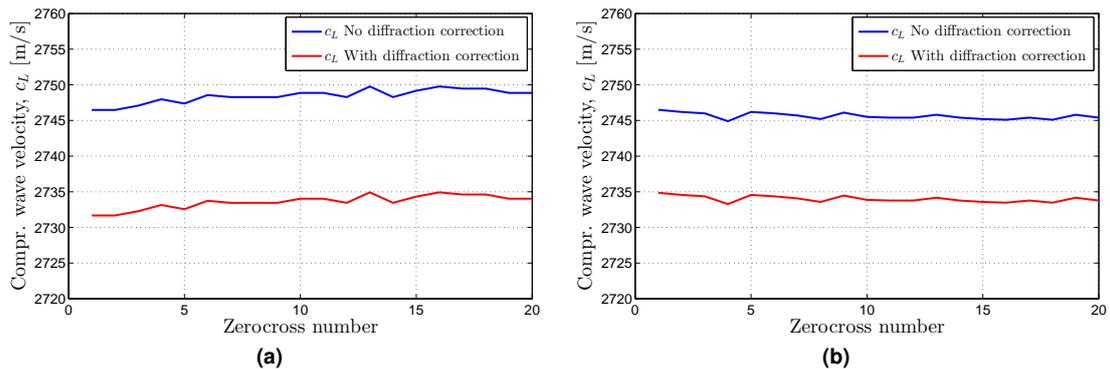

**Figure 13:** Compressional wave velocities for plexiglas specimen measured with the solid buffer method for the **a)** 20 mm specimen, **b)** 60 mm specimen.

The reproducibility of the compressional wave velocity measurements on the plexiglas specimen was ± 3 m/s for the immersion method and ± 5 m/s for the solid buffer method. The reproducibility was tested in the immersion method by changing the water in the immersion cell and demount and mount the adapter holding the specimen. For the solid buffer method, the reproducibility was tested by decoupling and cleaning the surfaces and adding new coupling fluid before reassembling the measurement setup. More than five measurements was done for both the immersion and solid buffer method for testing the reproducibility in the measurements.

## 5.2 Compressional wave velocity, Bentheim sandstone specimen.

Figure 14 shows the compressional wave velocity of the 20 mm Bentheim sandstone specimen measured using the immersion and the solid buffer methods. A 15 periods 500 kHz, 2 V input-signal is used on the signal generator. The receiving signal amplifier is set to 60 dB amplification. The porous sandstone is much more attenuating than the plexiglas and low-frequeny components in the signal might be present. To ensure that measurement A and measurement B are done in the steady-state region of the signal, a high-pass filter with cut-off frequency 300 kHz was used. The red and blue line show the results with and without diffraction correction applied. The dashed line in Figure 14b shows a measurement after the sandstone has been lying outside the water bath, draining for 5 minutes. The compressional wave velocity with diffraction correction is measured to be about 20 m/s lower with the solid buffer than using the immersion method.

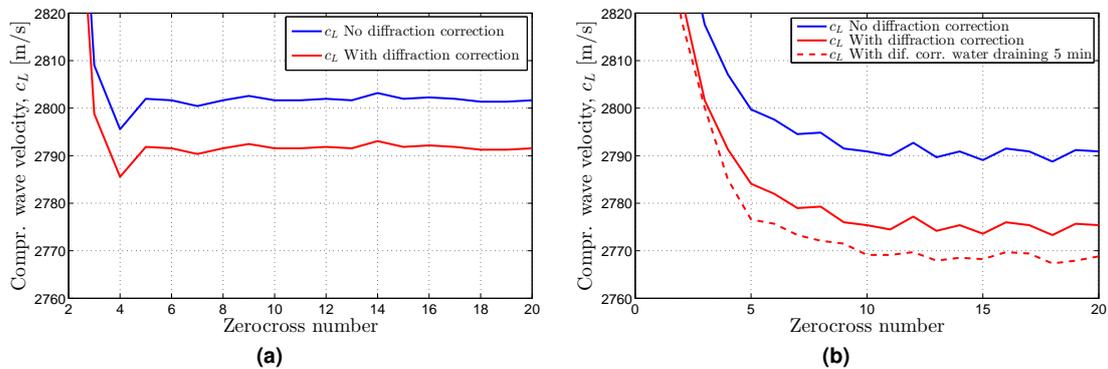

**Figure 14:** Compressional wave velocities for Bentheim sandstone disc with thickness 20 mm measured with **a)** the immersion method, **b)** the solid buffer method using 1 Nm torque.

In Figure 15 compressional wave velocity measurements are shown for the 47 mm long Bentheim sandstone specimen cut out from the same core sample as the 20 mm disc. Using the immersion method, the measured compressional wave velocity for the 47 mm cylinder is ∼ 45 m/s lower than the measured value for the 20 mm disc. Using the solid buffer method, the measured compressional wave velocity for the 47 mm cylinder is ∼ 15 m/s lower than the measured value for the 20 mm disc. While the measured compressional wave velocity for the 20 mm disc in Figure 14 is higher for the immersion method than the solid buffer method, this is not the case for the 47 mm long cylinder in Figure 15. For the 47 mm long cylinder, the compressional wave velocity is (∼ 15 m/s) lower when measuring with the immersion method compared to using solid buffer method with 1 Nm torque. Even this small amount of applied pressure affects the wave velocity measurements. When increasing the applied torque from 1 Nm to 3 Nm, the wave velocity increases almost 20 m/s for the 47 mm cylinder. One explanation for these deviations in the compressional wave velocity measurements might be due to cracks in the 47 mm cylinder. If cracks are present inside the rock, 1 Nm of applied torque might be enough to alter the stiffness of the Bentheim sandstone and hence the compressional wave velocity. The compressional wave velocity increases for a rock being subject to a hydrostatic pressure [13–17] and is also assumed to increase when an outer force is pressing on the specimen.

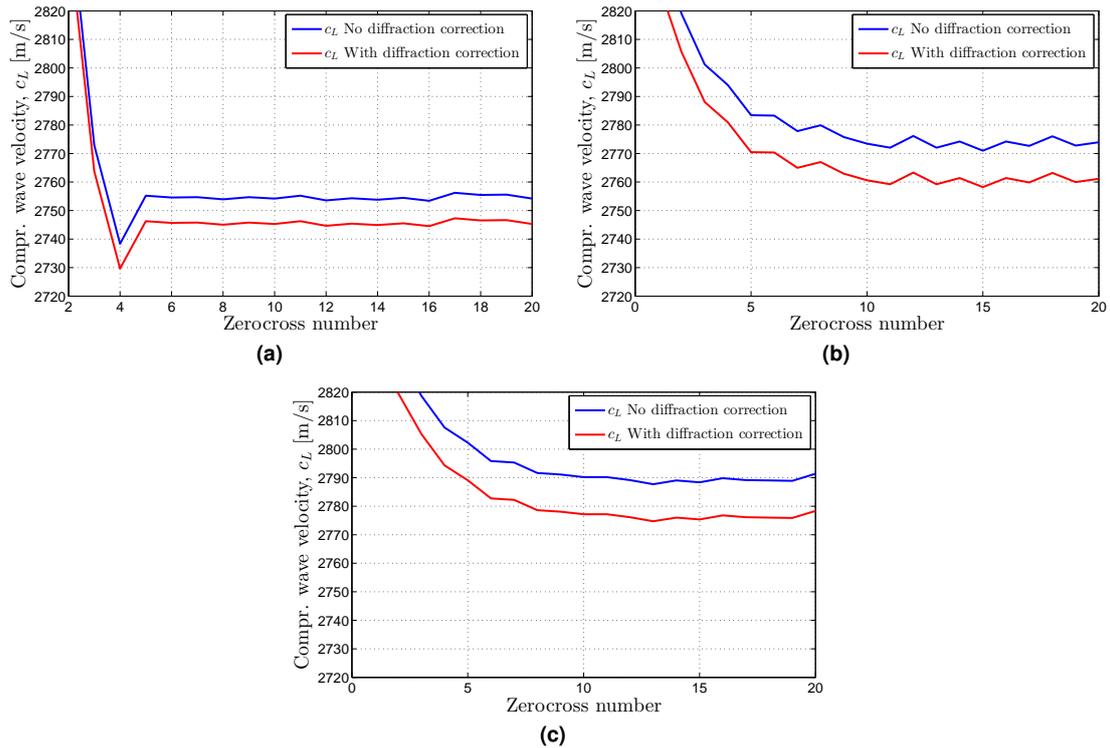

**Figure 15:** Compressional wave velocity for Bentheim sandstone cylinder with lenght 47 mm measured with **a)** the immersion method, **b)** the solid buffer method using 1 Nm torque, and **c)** the solid buffer method using 3 Nm torque.

The compressional wave velocity results for the 53 mm Bentheim cylinder are shown in Figure 16. The velocity is clearly higher in these measurements than in previous Bentheim sandstone measurements. This is probably due to the higher fluid saturation in the 53 mm specimen compared to the other Bentheim sandstone specimen. The change in compressional wave velocity from 1 Nm to 3 Nm is only 5 m/s and is a smaller change than for the 47 mm Bentheim sandstone cylinder in Figure 15. These measurements give a much better consistency between the immersion and solid buffer methods.

The reproducibility of the compressional wave velocity measurements on the 20 mm Bentheim sandstone specimen was found to be typically ±15 m/s while on the 47 mm and 53 mm Bentheim cylinders a reproducibility of ±10 m/s was obtained. These reproducibility-tests were found to be representable for both the immersion method and the solid buffer method. More than five measurements was done for both the immersion and solid buffer method for testing the reproducibility in the measurements.

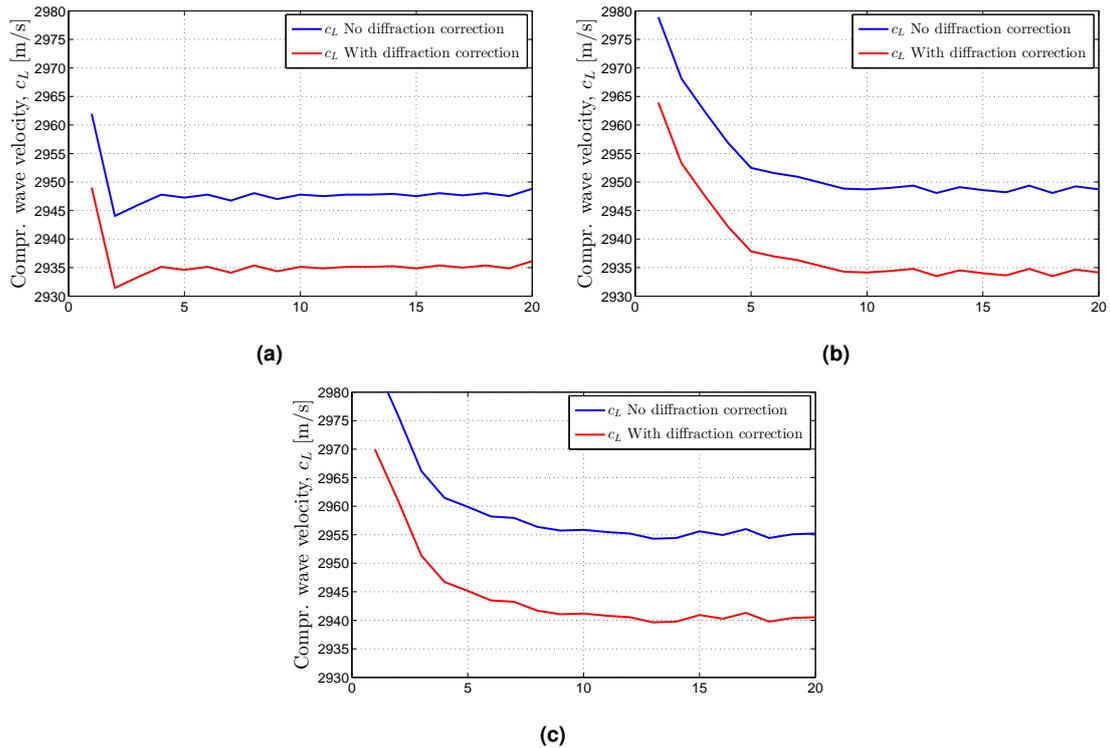

**Figure 16:** Compressional wave velocity for Bentheim sandstone cylinder with lenght 53 mm measured with **a)** the immersion method, **b)** the solid buffer method using 1 Nm torque, and **c)** the solid buffer method using 3 Nm torque.

## 6  Simulations

To investigate the influence on the solid buffer measurement method from possible sidewall reflections interfering with the measured signal, three different simulation configurations have been used. Configuration 1 represents the setup used in the measurements presented in section 5. The plexiglas buffers have a diameter of 35 mm and a length of 50 mm, see Figure 17. In configuration 2 the diameter of the buffers are increased from 35 mm to 115 mm, see Figure 18. In these simulations sidewall reflections will arrive later than in configuration 1. In configuration 3, the buffer diameter is 35 mm as in the measurements, but circumferential "rills" are introduced at the cylindrical surfaces, see Figure 19. These rills are introduced to see if roughened buffer-edges can help diminish sidewall reflections. The rills are 0.5 mm wide and 2 mm deep.

The voltage signals seen in Figure 17-19 is the measurement A of the respective configuration. The signals are obtained as described in section with 1 cycle, 500 kHz, 1 V input pulse on the signal generator. The signals are scaled to unity to easier compare the shape of the signals. Simulations of the compressional wave velocity of the plexiglas disc are shown in Figure 20-23. The red circles in Figure 17-18 highlight regions of interest. The first peak has a lower value in transducer configuration 1 than in transducer configuration 2. In configuration 1, a small ripple is seen at around 57 $\mu$s which is not so apparent in the wider buffer configuration. When carving in some rills in the buffers, shown in Figure 19, the scaled voltage signal differs even more to the wide buffer configuration in Figure 18 than without the rills.

The first reflections from the end-surfaces of the plexiglas buffers will arrive after 18 periods for frequencies 500 kHz. The first theoretical sidewall reflections in configuration 1 will arrive during the first period of the received signal. In configuration 2, the first theoretical sidewall reflections will arrive after 7 periods.

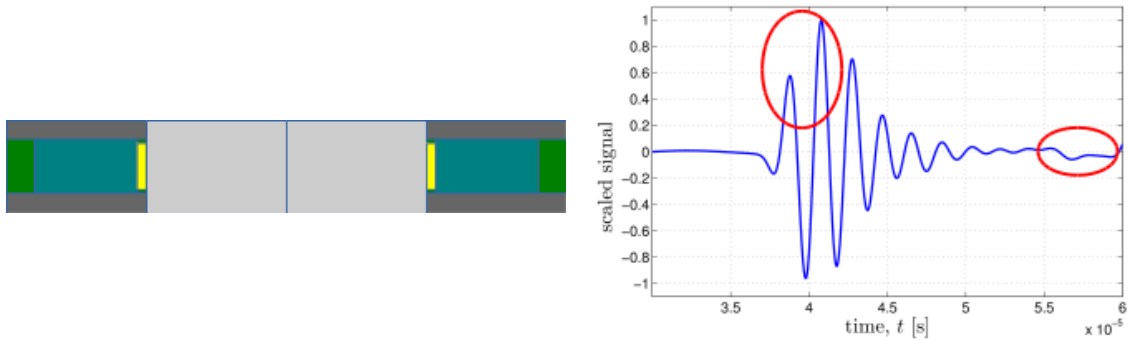

**Figure 17:** Transducer configuration 1 shown for measurement A and scaled simulated voltage signal, $V_2$.

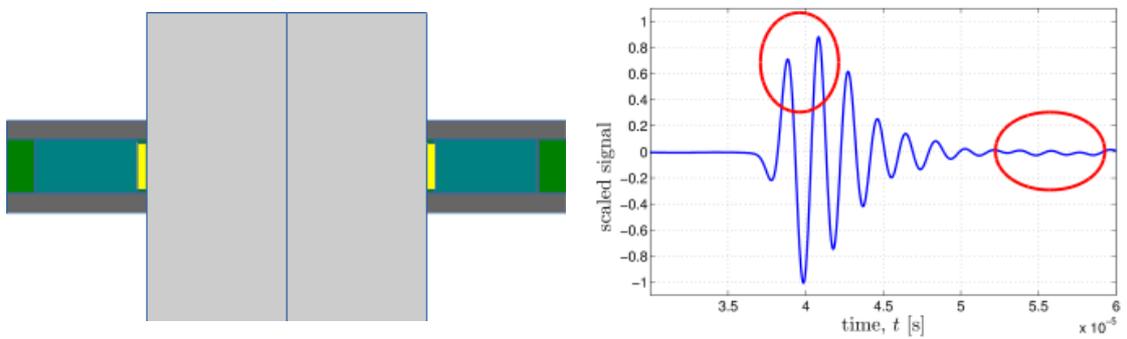

**Figure 18:** Transducer configuration 2 shown for measurement A and scaled simulated voltage signal, $V_2$.

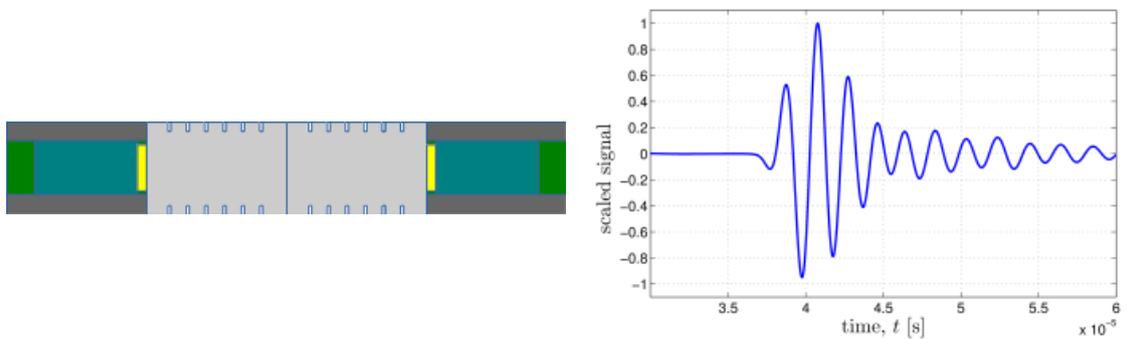

**Figure 19:** Transducer configuration 3 shown for measurement A and scaled simulated voltage signal, $V_2$.

Figure 20 shows more detailed simulations results for the solid buffer method with wide buffers (configuration 2). A plexiglas disc with thickness 20 mm and diameter 50 mm is used. The excitation signal, is a 15 cycles 500 kHz, 1 V signal burst as described in section 6. Figures 20a-20b show the transfer function, $V_2/V_1$. In Figures 20c-20d the transfer function $V_2/V_1$ is multiplied by the transfer function for the Thevenin circuit, $V_1/V_0$, and the spectrum of the 15 cycles 500 kHz, 1 V signal, $V_0$. The time domain signal in Figures 20e-20f are obtained by Fourier synthesis of the voltage spectra in Figures 20c-20d. The signal reaches steady state after approximately 4 periods and there are no clear signs of reflections in the signals.

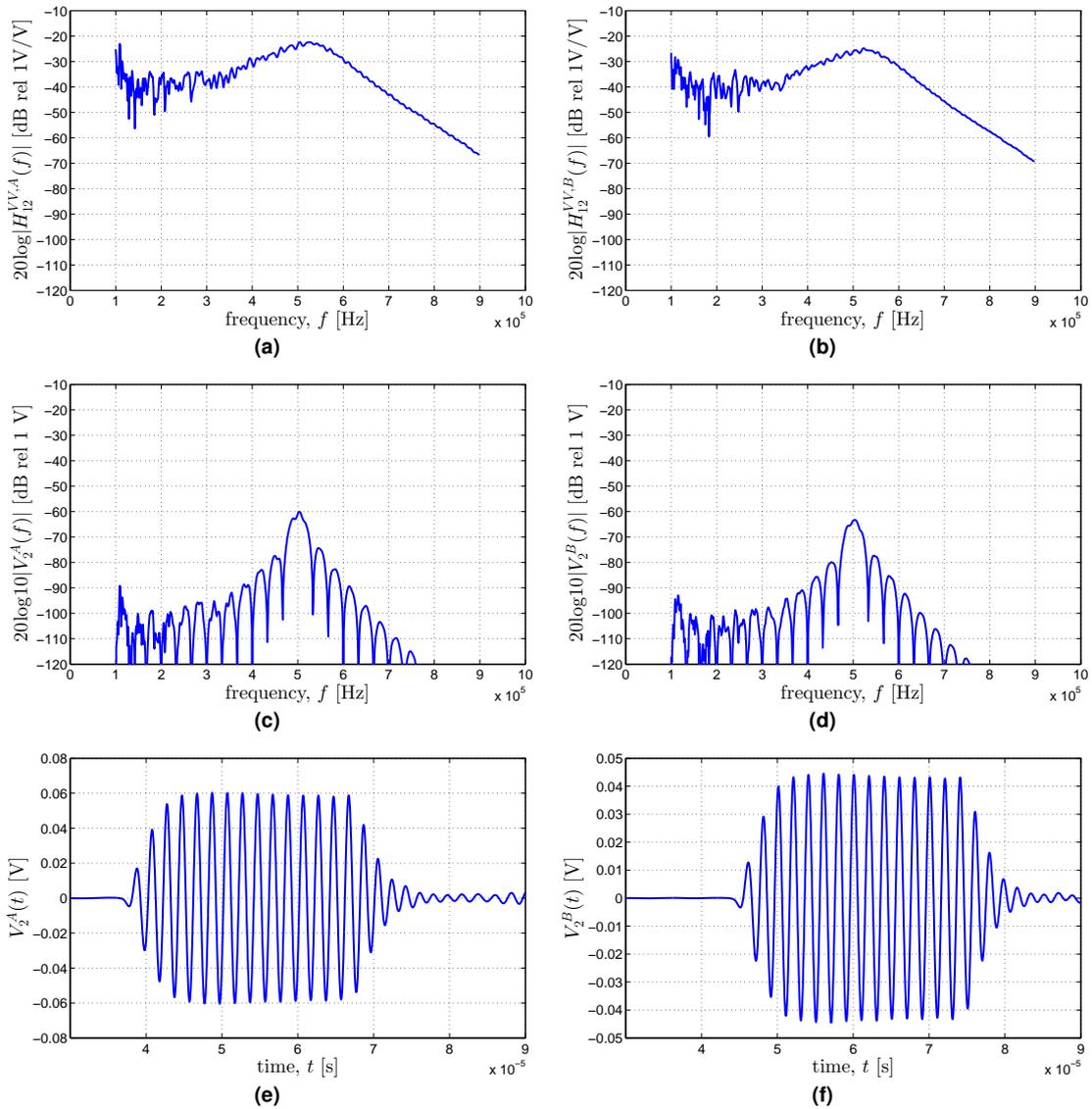

**Figure 20:** Simulated transfer functions and voltage signals using the solid buffer method with transducer configuration 2. **a)** Transfer function for "measurement A". **b)** Transfer function for "measurement B". **c)** Voltage spectrum, $V_2(f)$ for "measurement A". **d)** Voltage spectrum, , $V_2(f)$ for "measurement B". **e)** Time domain voltage signal, $V_2(t)$ for measurement A. **f)** Time domain voltage signal, $V_2(t)$ measurement for B.

The simulated compressional wave velocity of the plexiglas specimen using the wide buffer transducer configuration 2 is shown in Figure 21. Here it is seen that after diffraction correction, the estimated velocity is about 2696 m/s, which is 4 m/s below the specified input value, 2700 m/s. The zerocross technique as described in section 2 is used calculating the compressional wave velocity.

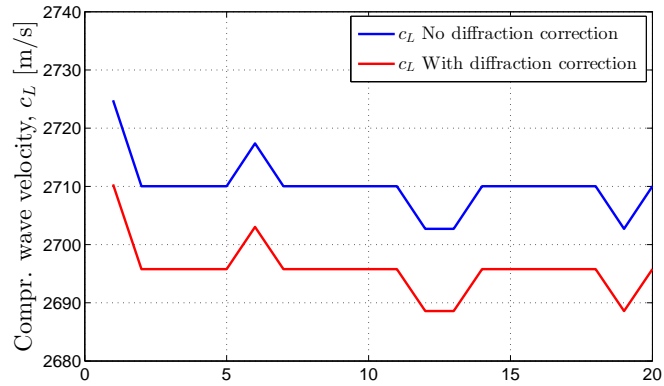

**Figure 21:** Simulated compressional wave velocity for a plexiglas disc with lenght 20 mm for the solid buffer method with transducer configuration 2.

In Fig 22, the time domain voltage signal of configuration 1 is shown. After approximately 10 periods there is an increase in the signal amplitude indicating sidewall reflections.

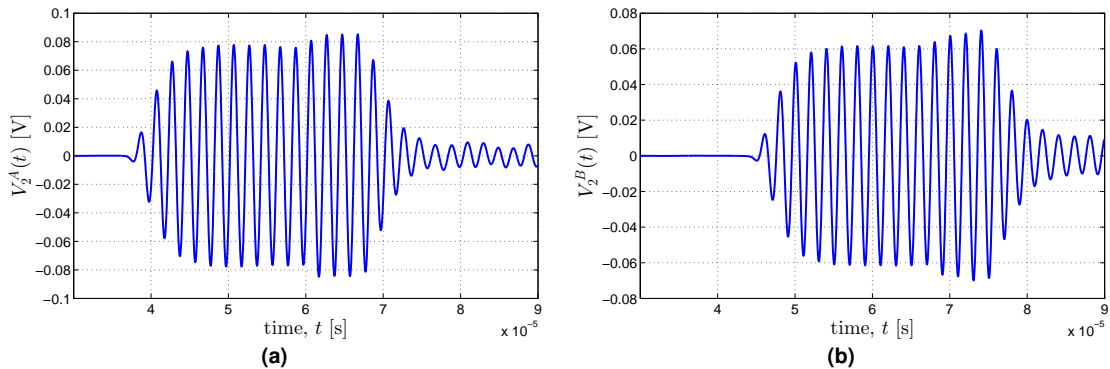

**Figure 22:** Simulated voltage signals for the solid buffer method for transducer configuration 1. **a)** Time domain voltage signal for measurement A. **b)** Time domain voltage signal for measurement B.

The simulated compressional wave velocity of the plexiglas specimen using the transducer configuration 1 is shown in Figure 21. A plexiglas disc with thickness 20 mm and diameter 50 mm is used. Here it is seen that after diffraction correction, the estimated velocity is about 2703 m/s, which is 3 m/s above the specified input value, 2700 m/s. This simulated value for the compressional wave velocity in plexiglas is about 7 m/s higher than the simulated value for transducer configuration 2. The zerocross technique as described in section 2 is used calculating the compressional wave velocity

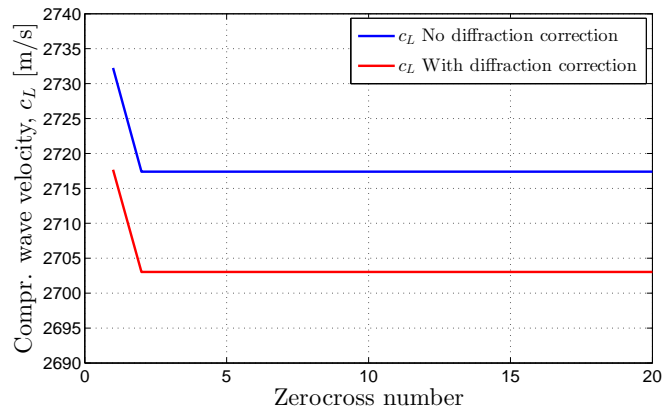

**Figure 23:** Simulated compressional wave velocity for a plexiglas disc with lenght 20 mm for the solid buffer method with transducer configuration 1.

## 7 Discussion

As described in section 1, a main objective of the present work is to evaluate the solid buffer method that is to be used with the pressure cell. Because of the limited space in the pressure cell, the solid buffer method is not an ideal method, since side-wall reflections might interfere with the waves propagating directly from the transmitter to the receiver.

The compressional wave velocities for the plexiglas specimen measured using the immersion and the solid buffer methods, are quite consistent which indicates that side-wall reflections have a minor effect on the measured compressional wave velocity. To further investigate how reflections act on the measurements, simulations of the solid buffer method was done, one simulation where the buffers were made wider than in the measurements (Figure 18) to avoid side-wall reflections, configuration 2, and one simulation with buffers as used in the measurements, configuration 1. The simulated compressional wave velocity for the configuration 1 (see Figure 23) is 5-10 m/s higher than the simulated compressional wave velocity in configuration 2 (see Figure 21).

To rely on these simulations the compressional wave velocity in the wide buffer configuration should be very close to 2700 m/s, which is the specified input compressional wave velocity. The simulated compressional wave velocities for the 20 mm plexiglas disc using configuration 1 and configuration 2, was 2703 m/s and 2696 m/s, respectively.

In this work, the nominal piezoelectric element radius of 12.5 mm has been used when calculating the diffraction correction. If using an effective piston radius of 10.55 mm, the simulated compressional wave velocity of the plexiglas after diffraction correction is 2699 m/s.

The effect of the sidewall reflections are seen when comparing Figure 18 with Figure 17 and 19. While no reflections are assumed in the relevant time window in configuration 2, it seems like reflections are present from the first period in the signal in configurations 1 and 3 (Figures 17 and 19, respectively). For the rilled buffers, side-wall reflections are dominating even more than in configuration 1. This is the main reason why buffers with roughened outer edges have not been used in the measurements.

The compressional wave velocity was measured on three different Bentheim sandstone specimen. For the 20 mm specimen, Figure 14, the wave velocity is measured to be about 15 m/s lower using the solid buffer method than with the immersion method. The wave velocity also drops 7-8 m/s after being out of the water bath for 5 min. These effects are not seen in the larger Bentheim cylinders in Figure 15-16. This can be due to water draining primarily out of the rocks at the outer edges and hence the smaller rock will be more affected by this effect. Especially the 47 mm specimen showed in Figure 15 shows another behavior than the 20 mm specimen. The compressional wave velocity increases from the immersion method to the solid buffer method with 1 Nm torque in the 47 mm specimen as compared to the 20 mm specimen. Visual inspection of the 47 mm Bentheim sandstone and the fact that these two specimen are cut out of the same core sample, gives reason to believe that there are some cracks present in the 47 mm cylinder. Cracks will lead to a lower compressional velocity and an outer pressure will lead to an increased wave velocity [53]. The 3 Nm torque exerted on the endcaps is observed to increase the compressional wave velocity far more in the 47 mm than in the 53 mm Bentheim cylinder. Other sources of uncertainties are expected to stem from taking the water saturated porous Bentheim sandstone up and down from the water bath, cleaning off, and putting on new coupling fluid. This treatment of the specimen can alter the saturation degree and hence the compressional wave velocity in the Bentheim sandstone specimen.

## 8  Conclusion

There seems to be several independent effects altering the measured compressional wave velocity using the solid buffer method in the pressure cell. These include side-wall reflections, diffraction effects, water draining from the saturated porous Bentheim sandstone, and the amount of torque excreted on the endcaps and hence an inward directed force on the specimen. Measurements indicate that draining effects are small for Bentheim sandstone cylinders.

For the 53 mm Bentheim sandstone cylinder, the immersion method can be used as a reference method for calculating the compressional wave velocity, i.e the same compressional wave velocities may be obtained for the immersion method and the solid buffer method when exerting a torque of 1 Nm on the endcaps. Cracks in the 47 mm Bentheim cylinder are probably the reason why the immersion seem unsuitable as a reference method for this specimen. Because of water draining from the 20 mm Bentheim porous sandstone disc, the immersion method and the solid buffer method give deviations in the compressional wave velocity of about 15 m/s. Deviations in compressional wave velocity measurements on the 47 mm Bentheim sandstone cylinder is believed to be due to cracks. The immersion and solid buffer methods give very good agreement when measuring the compressional wave velocity on the 53 mm Bentheim sandstone cylinder, which is believed to have no cracks of big magnitude. It is believed that longer porous sandstones are less prone to water draining.

Reflections from sidewalls will affect the compressional wave velocity measurements for the solid buffer method, but not critically. This effect seems to increase the measured wave velocity with 5-10 m/s in the simulations for the 20 mm plexiglas specimen. The diffraction correction calculated for a piston mounted in a rigid baffle of infinite extent will decrease the measured velocity by a magnitude of 10 m/s at 500 kHz.

The reproducibility of the compressional wave velocity measurements on the 20 mm Bentheim sandstone specimen was found to be typically ±15 m/s while on the 47 mm and 53 mm Bentheim cylinders a reproducibility of ±10 m/s was obtained. These reproducibility-tests were found to be representable for both the immersion method and the solid buffer method. More than five measurements was done for both the immersion and solid buffer method for testing the reproducibility in the measurements.

The solid buffer method is considered suitable for use in the pressure cell. Compressional wave velocity measurements have been done in the steady state region at frequency 500 kHz for various specimen made of plexiglas and Bentheim sandstone. The next step will be to monitor the compressional wave velocities while growing methane hydrate in the Bentheim sandstone.